\documentclass{elsart}
\usepackage{natbib}
\usepackage{graphicx}
\input epsf.sty
\begin{document}
\runauthor{A.Janiuk and B. Czerny}
\begin{frontmatter}
\title{Properties of the two-temperature corona model for active galactic
nuclei and galactic black holes}
\author[CAMK]{Agnieszka Janiuk\thanksref{kbn}}
\author[CAMK]{Bo\.zena Czerny\thanksref{kbn}}
\address[CAMK]{N. Copernicus Astronomical Center, Bartycka 18, 00-716 Warsaw, Poland}
\thanks[kbn]{Partially supported by the Polish State Committee for 
Scientific Research, grant 2P03D018.16}

\begin{abstract}
We study in detail the properties of the accreting corona model for active 
galactic nuclei and galactic black holes. In this model the fraction of the 
energy liberated in the corona at a given radius is calculated from the 
global parameters of the model (mass of the central object, accretion rate 
and viscosity parameter) and it appears to be a strong function of the radius.
The model predicts the relative decrease of the coronal hard X-ray emission 
with an increase of the accretion rate. The presented description of disc/corona interaction forms a basis for further studies of disc disruption mechanism.

\end{abstract}
\begin{keyword}
accretion, accretion discs -- black hole physics -- galaxies: active --
X-rays: galaxies -- X-rays: stars
\end{keyword}
\end{frontmatter}

\section{Introduction}

The accretion flow onto black holes both at the centers of active galactic
nuclei (AGN) and galactic black holes (GBH) consists of two phases: relatively
cold optically thick accretion disc responsible for multicolor black body
emission and a hot optically thin plasma responsible for a power law shape 
X-ray emission. 

There are two basic families of models of this two phase flow. 

The first family assumes the radial division of the flow into hot and cold 
part, with outer part cold and the inner part hot and optically thin. That
line of research originated with the classical paper of Shapiro, 
Lightman and Eardley (1976) followed by a number of papers which differed
with respect to the description of the heating and cooling mechanisms of the
hot part and the position of the radius dividing the hot from the cold
flow (e.g. Wandel \& Liang 1991, Narayan, Kato \& Honma 1997). Also a 
possibility of having an outer part hot and an inner part cold have been
recently discussed (Dullemond \& Turolla 1998).

The second family assumes the vertical division of the flow into hot and
cold part, i.e. describes an accretion disc with a hot corona. First papers
on this subject (e.g. Liang \& Price 1977, Bisnovatyi-Kogan \& Blinnikov 
1977) were further followed by a number 
of papers adopting
different descriptions of the fraction of energy liberated in the corona,
heating and cooling of the corona and the disc/corona interaction (e.g.
Haardt \& Maraschi 1991, Svensson \& Zdziarski 1994, \. Zycki, 
Collin-Souffrin \& Czerny 1995). 

Some phenomenological models actually filled this gap assuming that 
there can be
some overlapping between these two distinct possibilities, i.e. the innermost
part of the flow is hot and optically thin, an intermediate part consists
of a disc but covered by the hot plasma extending beyond the central region 
and the outermost part is a bare accretion disc (e.g. Poutanen, Krolik
\& Ryde 1997, Esin, McClintock \& Narayan 1997, Magdziarz \&
Blaes 1997). However, such an 
intermediate region did not result from
physically based assumptions but was introduced as a kind of ad hoc 
parameterization.

Both families of models have their advantages and disadvantages from the
observational point of view.

Models of a single compact hot plasma cloud surrounding the central black
hole are claimed to reproduce well the 'primary' X-ray spectrum for 
Cyg X-1 (e.g. Dove, Wilms \& Begelman 1997)
although other authors found some overlapping between the hot cloud  and cold
disc necessary (Poutanen et al. 1997) and perhaps stratification
in the hot plasma temperature is required (Ling et al. 1997, Gierli\' nski
et al. 1997). 
However, these models predict too compact X-ray emitting region in 
order to explain within a frame of pure Comptonization model (Kazanas, Hua \&
Titarchuk 1997)
long time/phase delays measured for that source (Cui et al. 1997). In the case
of AGN, their X-ray spectra are also well modeled by comptonization in a
single compact plasma cloud (e.g. Johnson et al. 1997 for NGC 4151, Gondek
et al. 1996 for composite Seyfert spectrum) and a reflection from the cold 
gas.

Corona models are equally successful in modeling the data if certain
arbitrary clumpiness of the corona is allowed (Haardt, Maraschi \& Ghisellini 
1994; see also Gondek et al. 1996).
Models assuming continuous corona over-produce the ratio of the 
hard X-ray emission to the optical/UV/soft X-ray component if the spectrum
is flat and the 
same ratio of the dissipation in the disc and in the corona is adopted at
all disc radii.

Both families of models usually contain a number of arbitrary parameters so 
they can give satisfactory fits to the observational data. This success makes 
the differentiation between the two geometrical models rather difficult.
Further progress can be made using two approaches: either to look for a
second order differences which may show up in high quality spectra or to
make a step back and incorporate into the model only well justified 
ingredients. 

This second, complementary approach is also valuable although it does not
provide immediately as precise description of the observational data as the 
phenomenological one. If we only allow for a well known or reasonably 
parameterized {\it physical} input into the model we are on a track to
prove or falsify the model. This approach requires first to study the
properties of the resulting model and its sensitivity to the principal 
ingredients. The next step is to study in detail the discrepancies between the
model predictions and the data in order to see whether those discrepancies 
can be accommodated within the frame of the model or not. This approach does 
not lead to the success (in the sense of reproducing data) immediately but
the ultimate goal of understanding the underlying physics of accretion makes
it worth to pursue.

A lot of work along such line is being done recently 
for advection-dominated accretion flow
(ADAF) solutions (e.g. Quataert \& Narayan 1998, Kato \& Nakamura 1998,
Gammie \& Popham 1998, Di Matteo et al. 1998).

In this paper we adopt the same approach but to a two-temperature corona 
model based on two 
assumptions: (1) the dissipation in the corona is proportional to the pressure
(2) vertical transition between the disk and the corona at each radius is 
determined by the marginal thermal stability of the Compton/atomic cooled
medium. Such a model does 
not require any
arbitrary parameterization of the transition between the hot and the cold 
plasma 
but nevertheless
reproduces a broad range of the ratios between the bolometric luminosity 
of the optically thick disc and hot plasma emission.  We show that this 
property results from strong radial dependence of the relative strength of the
corona which is the intrinsic property of the model.

The properties of the corona (i.e. the electron and ion temperature, the
optical depth and the fraction of total energy dissipated at a given
radius measured in Schwarzschild radii) only weakly depend on the central 
mass of 
the black hole if the accretion rate is measured in Eddington units. Therefore,
the same model applies without any major change to an AGN or a GBH. 
However, the
radiation spectra given by the model are significantly influenced by the
central mass due to the dependence of the disc temperature on its value.

The approach used is much simpler than that of \. Zycki et al. (1995) and
Witt et al. (1997). Instead, the attention is payed to the prediction of
the radial distribution of the disk/corona properties, the spectra and
the sensitivity of the model to the description of the physical processes 
involved.

The contents of the paper is as following. In Section~\ref{sect:struc} we
summarize the underlying assumptions which lead to much simpler description
than the full dynamical treatment of Witt et al. (1997). In 
Section~\ref{sect:spec} we describe the computation method of the spectra
integrated over the disk surface. Section~\ref{sect:ModelA} is devoted
to analysis of the radial properties of the simplest corona model 
and its predicted extension. In Sections~\ref{sect:Monte}, ~\ref{sect:nonloc},
and ~\ref{Coupling} we analyze the importance of the accuracy of
description of the X-ray transfer and ion-electron interaction. 
The results are discussed in 
Section~\ref{sect:diss} and conclusions given in Section~\ref{sect:conc}.



\section{Model}
\label{sect:model}

\subsection{Disc/corona structure}
\label{sect:struc}

In this model of accretion flow onto a black hole we assume that the 
(stationary) accretion is ultimately responsible for the emission of the 
radiation and therefore the accretion rate, $\dot M$, together with the mass
of the central black hole, $M$, are the basic parameters of our model. 
The accretion proceeds predominantly through the disk which is surrounded,
wherever possible, by a hot optically thin corona.

We assume that the hot optically thin corona is a two-temperature medium,
as in the paper of Shapiro, Lightman
\& Eardley (1976), i.e. the ion temperature is higher than the electron
temperature, the loss of gravitational energy by accreting gas is transported
directly to the ions, Coulomb
coupling transfers this energy to electrons and finally electrons cool down
by Inverse Compton process, with disc emission acting as a source of soft
radiation flux.

We approximate the distribution of the angular momentum in the disk and the 
corona by the Keplerian distribution so the sum of the flux generated by the 
disk, $F_d$, and by the corona, $F_c$, at any given radius is determined 
by the standard formula (see e.g. Kato, Fukue \& Mineshige 1998)
\begin{equation}
F_d +F_c = {3 G M \dot{M} \over 8 \pi r^3} f(r),
\end{equation}
where $f(r)$ represents the Newtonian boundary condition at the margi\-nal\-ly
stable orbit
\begin{equation}
f(r)= 1- (3R_{\rm Schw}/r)^{1/2}.
\end{equation}

We describe the local heating of the ions  in terms of the
$\alpha$ viscosity introduced by Shakura \& Sunyaev (1973) and we neglect the
coronal radiation pressure since it should not contribute to energy generation
in optically thin medium:
\begin{equation}
F_c =  {3\over 2} \Omega_K \alpha P_o H {\sqrt {\pi \over 2}} 
\end{equation}
where $P_o$ is the pressure at the basis of the corona, $H$ is the pressure  
scale height of the corona
given by the ion temperature, $T_i$, under the assumption of the hydrostatic 
equilibrium
\begin{equation}
H = {1 \over \Omega_{K}}{\sqrt{kT_i \over m_H}}
\end{equation}
and the factor ${\sqrt \pi /2}$ results from the vertical integration of the
pressure (Witt, Czerny \& \. Zycki 1997, hereafter WCZ, Appendix D).

Such a formulation is independent from the physical mechanism of the corona
heating since the scaling with pressure may correspond either to accreting
corona which is powered directly by the release of the fraction of the 
gravitational energy of the accreting gas within the corona itself (e.g.
\. Zycki, Collin-Souffrin \& Czerny 1995, WCZ) or to the 
magnetic heating (e.g. Svensson \& Zdziarski 1994).

The net flux, $F_c + F_{nl}$, is subsequently transported to electrons
through  the electron-ion Coulomb 
interaction, as  described by the following equation (Shapiro,
Lightman \& Eardley 1976)  

\begin{equation}
F_c + F_{nl} ={3\over 2}  {k (T_i -T_e) \over m_H}
\left[ 1 + \left( {4 k T_e\over m_e c^2}\right)^{1/2} \right]
\nu_{ei} \rho_o H {\sqrt \pi \over 2}, 
\end{equation}
where $T_e$ is the electron temperature, and
\begin{equation}
\nu_{ei}= 2.44 \times 10^{21} \rho_o T_e^{-1.5}\, \ln \Lambda ~~~[s^{-1}];
\quad {\rm with} \quad \ln \Lambda \approx 20
\end{equation}
is the electron-ion coupling rate. Here the density $\rho_o$ is related
to the corona thickness and its optical depth $\tau_{es}$
\begin{equation}
\tau_{es} = k_{es} \rho_o H.
\end{equation}
We discuss the accuracy of this description and the dependence of the model
on the ion-electron coupling rate in Section~\ref{Coupling}.

The corona is assumed to be isothermal and its vertical density profile 
is taken into account only through the numerical factors in Eqs. (3) 
and (5).

The energy flux transferred to electrons has to be carried away by soft
X-ray photons scattered by the corona 
\begin{equation}
 F_c + F_{nl} = A(\tau_{es},T_e,T_s) F_{soft}
\end{equation}
which constitute hard X-ray coronal emission. We discuss the results for two
descriptions of the Compton amplification factor $A$. In 
Section~\ref{sect:ModelA} we 
present the
results obtained using a simple analytic description of this factor by the
Compton parameter $y$  as in WCZ) 
$$ A(\tau_{es},T_e,T_s) = e^y - 1 ;   y=\tau_{es} { 4 k T_e \over m_ec^2}
 \left( 1 + {4 k T_e\over m_e c^2} \right) \hfill (8a) $$\noindent
since is shows the properties of the model used by Czerny, Witt \&
\. Zycki (1997) and Kuraszkiewicz, Loska \& Czerny (1997) 
to compare to the AGN data. This method is very approximate
but simple and convenient since it is 
independent on the soft photon energy and in that case the corona model is
independent on the mass of the central black hole and can easily give results
appropriate for a
qualitative discussion (see WCZ, 
in particular Appendix C). 
In Section~\ref{sect:Monte} we compare those results with the results based on the 
Monte Carlo 
simulations
of the Comptonization process. The Monte Carlo code employs the method 
described by Pozdnyakov, Sobol \& 
Sunyaev (1983) and G\`{o}recki \& Wilczewski (1984). Assuming slab geometry
(Thomson optical depth $\tau_{es}$ and electron temperature $T_e$)
and the soft photons spectrum as a black body of temperature $T_s$ determined
by $F_{soft}$, we compute the amplification factor
$A$ on a grid of $T_e,\ \tau_{es},\ {\rm and}\,T_s$. We compute $A$ at each
radius by interpolation. This method is more accurate than the use of the
Compton parameter $y$. It introduces
the dependence of the model on the mass of the central black hole 
(through $T_s$) and allows for more quantitative discussion.  

The soft flux from the formula (8) is determined by the 
radiative coupling between the disc and the corona, as  described in the
basic paper of Haardt and Maraschi (1991). It means that the soft flux
emitted by the disk consists of the sum of the flux generated by the disk 
interior and the fraction of the coronal radiation absorbed by the disk
\begin{equation}
F_{soft} = F_d + \eta (F_c + F_{nl})(1 - a)
\end{equation}
In our basic model we assume that the fraction $\eta$ 
of the coronal flux directed
towards the disk is equal 0.5 and 
the energy averaged albedo $a$ is equal 0.2.

Our system of equations determining the corona structure, including the
fraction of the energy, $f$, liberated in the corona ($f = F_c /(F_c + F_d)$)
is closed by the following considerations.

The vertical division of the medium into hot corona and cold accretion disc 
at every radius should not be arbitrary, as it is customary assumed.
Indeed, such a division results naturally 
from the criterion of
thermal instability in an irradiated medium studied by Krolik, McKee and
Tarter (1981). The ionization stage of the medium is conveniently expressed
through a ionization parameter $\Xi$ defined as a ratio of the ionization 
radiation pressure to the gas pressure
\begin{equation}
\Xi = {\eta (F_c + F_{nl}) \over c P_o}. 
\end{equation}
The transition from cold to hot medium is characterized 
quantitatively by a specific value of the ionization parameter $\Xi$ which 
scales with the electron temperature as
\begin{equation}
\Xi = 0.65 (T_e/10^8 {\rm K})^{-3/2}
\end{equation}
(Begelman, McKee \& Shields 1983; see also \. Zycki et al. 1995). We discuss
this scaling in Section~\ref{Coupling}.

The set of equations (1) - (11) allows to calculate the radial structure 
of the corona as a function of global model parameters, i.e. mass  
of the black hole, $M$, the accretion rate, $\dot M$, and the viscosity
parameter $\alpha$ in the corona, if the nonlocal contribution $F_{nl}$
is specified. We frequently use throughout the paper the dimensionless 
accretion rate 
$\dot m$ 
measured in the Eddington units. 
\begin{equation}
\dot M_{Edd}=3.52  { M \over 10^8 M_{\odot}} [M_{\odot}/yr],
\end{equation}
where $M$ is the mass of the central black hole. The value of the mass does
not influence strongly the radial dependence of the coronal temperatures and 
its 
optical depth but it scales with
the absolute value of the corona luminosity. We assume a non-rotating black 
hole so the inner disc radius is located at 3$R_{Schw}$ and critical 
accretion rate was defined using the pseudo-Newtonian efficiency of accretion
equal to 1/16 and pure hydrogen opacity.

The computations of the disc structure are not required to complete the model.
Therefore, for the purpose of this study, we do not have to specify whether
the corona is actually an accreting corona transporting the mass and angular 
momentum 
in the same way as the disc does or it is coupled to the disc through the
magnetic field and the disc has to carry the entire angular momentum flux
(as assumed e.g. by Svensson \& Zdziarski 1994) whilst the energy generation
is proportional to the coronal gas pressure.
However, having the corona properties determined, we can also study the
vertical structure of the disc with appropriate boundary conditions at the
disc surface (R\' o\. za\' nska, et al. 1999) and assumption about the
angular momentum transport.

The presented model differs from the model studied by WCZ by being simpler
and at the same time more general. This model does not contain any dynamical
terms connected with the motion of the coronal gas  and therefore allows 
both for a slow inflow or outflow of the corona. We also study in
detail the importance of the exact description of the Compton amplification
factor and discuss the problem of corona formation in the context of the
radiation pressure instability within a standard disc.

\subsection{Spectra}
\label{sect:spec}

We calculate the radiation spectrum emitted at each radius $r$ separately.
 
Emission from the optically thick disc is computed neglecting the bound-free
transitions but taking into account the effect of electron scattering
(Czerny \& Elvis 1987). The local density at the disc surface is determined
by the hydrostatic equilibrium with the corona if corona develops at that radius, or it is taken
to be equal to 0.1 of the mean disc density in the absence of the corona at
that radius. This simplified method give similar results to the more advanced
method developed for a disc without a corona by D\" orrer et al. (1996). 

Soft photons emitted locally by the disc are  
Comptonized by the corona if a corona develops
at that radius. Its parameters:
the optical depth $\tau_{es}(r)$ and the electron temperature $T_e(r)$ are
computed from the model. The effect of Comptonization in our Model A 
is calculated as
by Czerny \& Zbyszewska (1991). In the case of Model B Monte Carlo simulations
are performed, as described by Janiuk, \. Zycki \& Czerny (1999).

At present we do not include the spectral component which results from the 
reflection of the X-rays from the disc surface since we concentrate on the
qualitative study of the model possibilities. Any future attempt to fit the
model to the observational data will have to include this component as well.

The final disc spectrum is computed by integration over the disc surface
assuming an inclination angle equal zero (i.e. top view). 
All computations are
done for a non-rotating black hole and the relativistic corrections were
neglected.

\section{Results}
\label{sect:res}

\subsection{Model A: simplified description of Comptonization}
\label{sect:ModelA}

In this section we present the properties of the corona calculated without
any nonlocal heating-cooling term, i.e. $F_{nl}=0$ in Equations (1)-(11). We
describe the efficiency of the Compton cooling using the analytic
approximation (Eq. 8a). Such a corona is fully determined at each radius 
by just two
parameters: the accretion
rate $\dot m$ expressed in the Eddington units and the viscosity parameter
$\alpha$, if the radius is expressed in units of the Schwarschild radii.
Such a scaling is natural if a disc temperature does not have to be
considered and it was frequently used to reduce the number of free parameters
in the description of a hot plasma (see e.g. Bj\"ornsson \& Svensson 1992).   

\subsubsection{The relative strength of the corona}
\label{sub:xi}

The most important property of the accreting corona model is the strong 
radial dependence of the strength of the corona.

\begin{figure}
\epsfxsize = 80 mm \epsfbox[50 180 560 660]{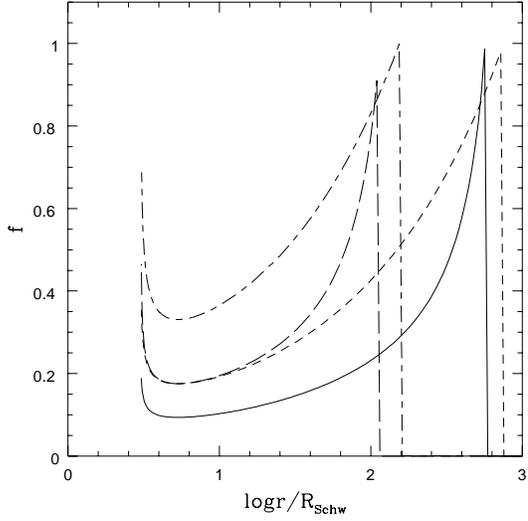}
\caption{Model A: The fraction of the energy dissipated in the corona as a function of
radius for two values of the viscosity parameter $\alpha$ and two values of
the accretion rate $\dot m$: $\dot m = 0.1$, $\alpha= 0.03$ (short dashed line),
 $\dot m = 0.1$, $\alpha= 0.33$ (continuous line),  $\dot m = 0.01$, 
$\alpha= 0.03$ (short - long 
dashed) and  $\dot m = 0.01$, $\alpha= 0.33$ (long dashed)
\label{fig:f}}
\end{figure}

An example of the radial dependence of the fraction of energy generated in the
corona is shown in Figure~\ref{fig:f}. The local value $f$  
is defined (after Haardt \& Maraschi 1991) 
as the fraction of energy dissipated in the corona. Note that WCZ 
used $\xi$ - a fraction of energy liberated in the disc - but these
two quantities are complementary (i.e. $f=1 - \xi$).

Simplified analytical solutions to the corona structure show (see
WCZ, Appendix C) that approximately this fraction of the energy $f$ 
increases with the radius $r$ as $f \propto r^{3/8}$. It means that the
X-ray flux $F_X(r)$ actually decreases with radius since the total flux 
generated in the disc decreases as $r^{-3}$ but this decrease is
slower than the decrease of the total flux: $F_X(r) \propto  r^{-3+3/8}$.
This is important from the point of view of the spectral shape of the
reflection component and the iron $K_{\alpha}$ line but, at present, we 
do not include the reflection component in the spectra.

The corona covers only the inner part of the disc starting from a certain 
radius $r_{max}$. At that radius all the energy is liberated in the corona
($f$=1). At larger radii no corona
solutions of our equations exist. The Compton cooling provided by the disc
is too large under the adopted assumptions about the corona structure. Any
additional (constant) cooling decreases $r_{max}$ whilst any additional
(constant) heating increases it.

\begin{figure}
\epsfxsize = 80 mm \epsfbox[50 180 560 660]{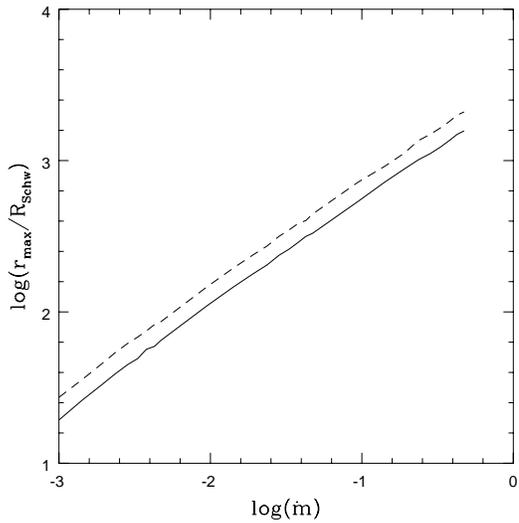}
\caption{Model A: The dependence of the extension of the corona on the accretion rate
for the viscosity parameter $\alpha$ equal 0.03 (dashed line) and
0.33 (continuous line)
\label{fig:ext}}
\end{figure}

Therefore there is a strong and 
discontinuous change of accretion flow structure at that radius since for
larger radii all the accretion proceeds thorough a disc whilst suddenly,
at $r_{max}$, rapid evaporation takes
place, the dissipation of energy concentrates in the hot corona and the cold 
disc is heated up only by X-ray illumination. Closer in the relative strength of the
corona decreases. 

The dependence of the radial extension $r_{max}$ of the corona on 
the accretion
rate is shown in Figure~\ref{fig:ext}. The size of the corona is considerable, 
covering the disc up to $\sim 1 000 R_{Schw}$ for large accretion rate but
it decreases significantly for smaller accretion rates, down to 
about $100 R_{Schw}$ for sources radiating at 1\% of the Eddington 
luminosity. The influence of the value of the viscosity parameter 
on the extension of the corona is rather weak.

\subsubsection{Ion and electron temperatures and the optical depth}

The ion temperature is a decreasing function of the radius (see Figure~\ref{fig:ion} ). 
It falls down almost linearly (see simplified formulae in Appendix C of WCZ).

In this paper we use simplified description of the vertical hydrostatic
equilibrium and we have to check afterwards whether the solution can 
actually be in the hydrostatic equilibrium. The approximate criterion 
is that the ion temperature should be smaller than the local virial
temperature. We see from Figure~\ref{fig:ion} that for low values of the viscosity
parameter this requirement is not satisfied unless the accretion rate
is smaller than about 1\% of the Eddington value. In WCZ the vertical
structure was calculated taking into account the transonic 
vertical outflow from the corona. In that case the solutions for
small values of viscosity parameter and large accretion rates simply
did not exist (see Figs. 1 and 3 of WCZ).

The pressure height scale of the corona defined as
\begin{equation}
H_P=({ k T_i r^3 \over GM m_H})^{1/2}
\end{equation}
increases almost linearly with the radius (see Figure~\ref{fig:h}) and the $H_P/r$ ratio is almost
constant.

\begin{figure}
\epsfxsize = 80 mm \epsfbox[50 180 560 660]{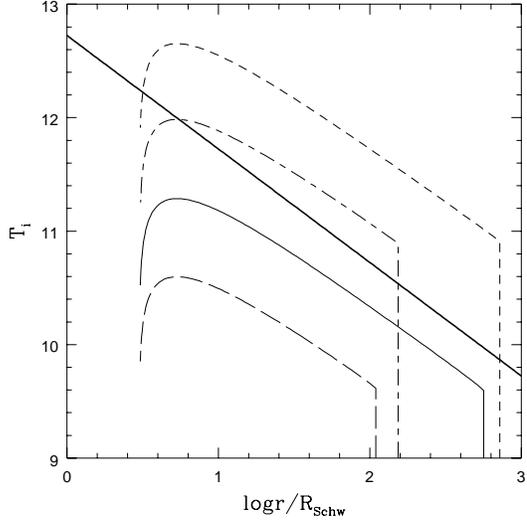}
\caption{Model A: The ion temperature as a function of
radius for two values of the viscosity parameter $\alpha$ and two values of
the accretion rate $\dot m$: $\dot m = 0.1$, $\alpha= 0.03$ (short dashed line),
 $\dot m = 0.1$, $\alpha= 0.33$ (continuous line),  $\dot m = 0.01$, 
$\alpha= 0.03$ (short - long 
dashed) and  $\dot m = 0.01$, $\alpha= 0.33$ (long dashed). The 
thick straight line shows the distribution of the local 
virial temperature.
\label{fig:ion}}
\end{figure}

\begin{figure}
\epsfxsize = 80 mm \epsfbox[50 180 560 660]{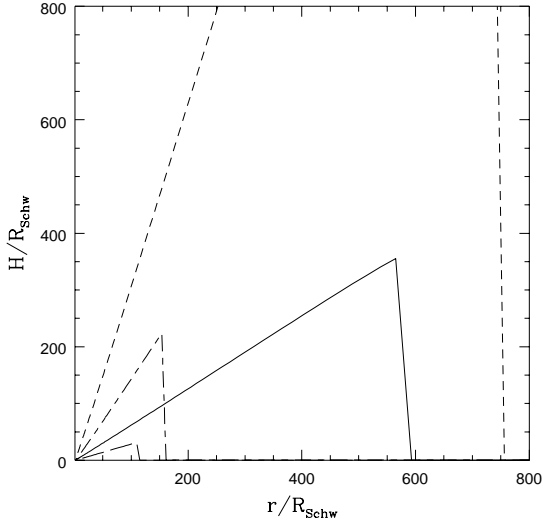}
\caption{Model A: The approximate shape of the corona given as a pressure height scale
for two values of the viscosity parameter $\alpha$ and two values of
the accretion rate $\dot m$: $\dot m = 0.1$, $\alpha= 0.03$ (short dashed line),
 $\dot m = 0.01$, $\alpha= 0.33$ (continuous line),  $\dot m = 0.01$, 
$\alpha= 0.03$ (short - long 
dashed) and  $\dot m = 0.01$, $\alpha= 0.33$ (long dashed)
\label{fig:h}}
\end{figure}

The formal solutions for high accretion rate and low viscosity which
are characterized by the ion temperature being higher than the virial
temperature show their problems in  Figure~\ref{fig:h} as well since in those cases
the height of the corona at a given radius is larger than that radius.
Therefore our model does not offer correct solution beyond certain 
range of parameters. If the model includes all the dynamical terms as in WCZ 
the physical problem of the character of the accretion flow is not solved since
the solutions of the full set of differential equations (which require 
hydrostatic equilibrium at the basis of the corona) disappear when the ion 
temperature would exceed the virial temperature. 

Since the density in the corona decreases outward the efficiency of Coulomb 
interaction 
between the ions and electrons decreases. However, both the disc and the 
corona bolometric luminosity go rapidly down with the radius. 
Therefore the electron temperature rises
outwards and the ratio of the ion to electron temperature decreases outwards.
The highest value of the electron temperature is reached at $r_{max}$.
Its value is roughly of the order of $3 \times 10^9$ K, 
independently from 
the accretion rate and the viscosity $\alpha$. The precise value varies
from $1.5 \times 10^9$ K for $\alpha$ equal 0.33 to $4.2 \times 10^9$ K
for $\alpha =0.03$ and its dependence on accretion rate is negligible.
Such a universal value is an interesting
property of our model and it is directly related to the predicted
properties of the spectra.

\begin{figure}
\epsfxsize = 80 mm \epsfbox[50 180 560 660]{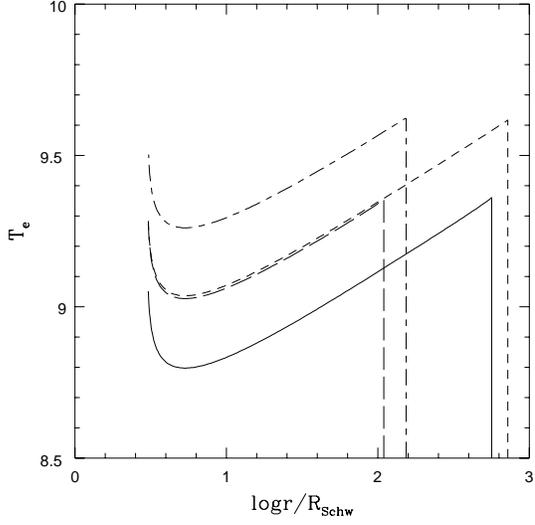}
\caption{Model A: The electron temperature as a function of
radius for two values of the viscosity parameter $\alpha$ and two values of
the accretion rate $\dot m$: $\dot m = 0.1$, $\alpha= 0.03$ (short dashed line),
 $\dot m = 0.1$, $\alpha= 0.33$ (continuous line),  $\dot m = 0.01$, 
$\alpha= 0.03$ (short - long 
dashed) and  $\dot m = 0.01$, $\alpha= 0.33$ (long dashed)
\label{fig:electron}}
\end{figure}

\begin{figure}
\epsfxsize = 80 mm \epsfbox[50 180 560 660]{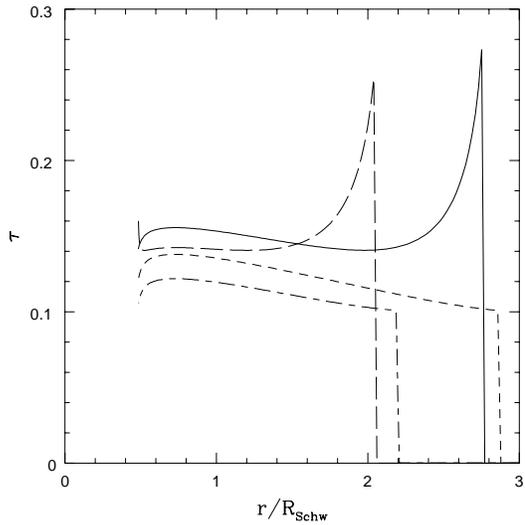}
\caption{Model A: The optical depth as a function of
radius for two values of the viscosity parameter $\alpha$ and two values of
the accretion rate $\dot m$: $\dot m = 0.1$, $\alpha= 0.03$ (short dashed line),
 $\dot m = 0.1$, $\alpha= 0.33$ (continuous line),  $\dot m = 0.01$, 
$\alpha= 0.03$ (short - long 
dashed) and  $\dot m = 0.01$, $\alpha= 0.33$ (long dashed)
\label{fig:depth}}
\end{figure}

\subsubsection{Radiation spectra for AGN}
\label{sub:speAGN}

Standard optically thick accretion disc models for parameters 
adequate for AGN predict that most emission concentrates in UV/soft
X-ray band forming a Big Blue Bump spectral component. In our model
this basic property is preserved 
but a fraction of the total bolometric luminosity
is emitted also in hard X-rays due to the Compton cooling of the 
corona.

\begin{figure}
\epsfxsize = 80 mm \epsfbox[50 180 560 660]{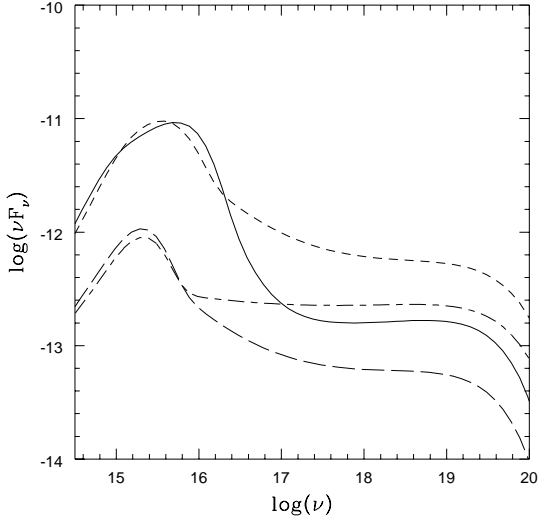}
\caption{Model A: The radiation spectrum for $M=10^8 M_{\odot}$ and for two values of the viscosity parameter $\alpha$ and two values of
the accretion rate $\dot m$: $\dot m = 0.1$, $\alpha= 0.03$ (short dashed line),
 $\dot m = 0.1$, $\alpha= 0.33$ (continuous line),  $\dot m = 0.01$, 
$\alpha= 0.03$ (short - long 
dashed) and  $\dot m = 0.01$, $\alpha= 0.33$ (long dashed)
\label{fig:spAGN} }
\end{figure}

The fraction of the bolometric luminosity emitted in the form of 
hard X-rays decreases with increasing accretion rate. It also
strongly depends on the adopted viscosity parameter, as seen from Figure~\ref{fig:spAGN}. Larger viscosity leads to more profound Big Blue Bump since
the radial extension of the corona is larger in that case and the
generation of hard X-rays is systematically shifted towards larger
disc radii where the gravitational energy available is lower.

Hard X-ray band above 20 keV is dominated by the emission of the
outermost part of the corona. Since the temperature there is high
and almost universal the spectrum is always practically 
flat on $\nu F_{\nu}$
diagram and its extension is only weakly dependent on the
viscosity parameter  and almost independent on the accretion rate.

In the soft X-ray band the spectrum shows more curvature than
the simple models composed from a standard accretion disc without
a corona and a single power law since the X-ray emission generated at
inner radii provide an additional steeper component since the
corona temperature there is lower than in the outer parts.   

Generally, the spectra for massive black holes are easy to parameterize
since the contribution from the disc and the corona form quite well separated
components. From the observational point of view they can be well represented
giving the $\alpha_{ox}$ index (which expresses directly the dominance of the
Big Blue Bump) and the slope of the hard X-ray part, $\alpha_x$.

The distribution of the  $\alpha_{ox}$ was studied by Czerny et al. (1997).
It compares favorably with the observed distribution for quasars and Seyfert
galaxies although there are a few objects with properties beyond
the range expected from the model.

\begin{figure}
\epsfxsize = 80 mm \epsfbox[50 180 560 660]{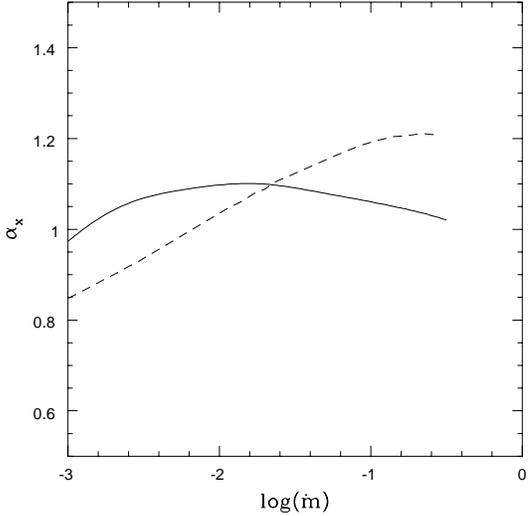}
\caption{Model A: The dependence of the 2-10 keV energy index $\alpha_x$ 
on the accretion rate for $M=10^8 M_{\odot}$ and the viscosity parameter 
$\alpha$ equal 0.03 (dashed line) and 0.33 (continuous line)
\label{fig:indexAGN}}
\end{figure}

The dependence of $\alpha_x$ on the accretion rate and the viscosity parameter
$\alpha$ is shown in Figure~\ref{fig:indexAGN}. 

We see that there is a weak trend in the change of the hard X-ray slope with
the accretion rate. In the case of low viscosity the spectra are 
systematically harder for lower
luminosity to the Eddington luminosity ratio. For high viscosity, this 
dependence is not monotonic.

\subsubsection{Radiation spectra for GBH}
\label{sub:speGBH}

As in the standard accretion disc models, low value of the mass of the central 
black hole results in much higher disc temperature and the optically thick
component peaks in the soft X-ray band. This fact affects the spectral
shape even in hard X-rays although the physical parameters of the corona
model like the electron temperature and the optical depth are independent on 
the mass of the central black hole. Since the hard X-ray emission is caused
by Comptonization of the optically thick disc emission higher disc temperature
changes the slope of the hard X-rays leading to spectra being 
systematically steeper (softer) than in the case of AGN.

\begin{figure}
\epsfxsize = 80 mm \epsfbox[50 180 560 660]{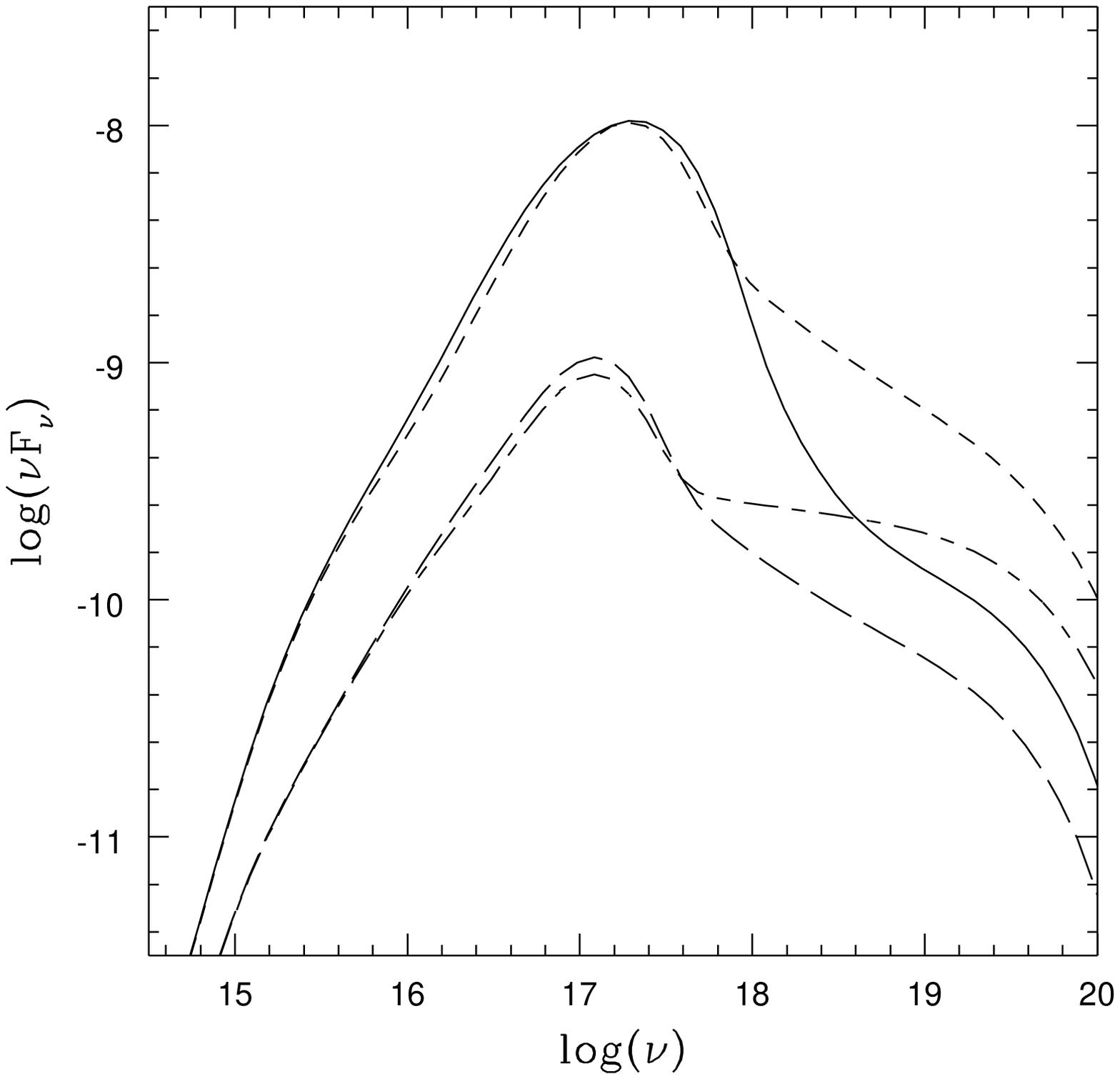}
\caption{Model A: The radiation spectrum for $M=10 M_{\odot}$ and for two values of the viscosity parameter $\alpha$ and two values of
the accretion rate $\dot m$: $\dot m = 0.1$, $\alpha= 0.03$ (short dashed line), $\dot m = 0.1$, $\alpha= 0.33$ (continuous line),  $\dot m = 0.01$, 
$\alpha= 0.03 $ (short - long 
dashed) and  $\dot m = 0.01$, $\alpha= 0.33$ (long dashed)
\label{fig:spGBH}}
\end{figure}

\begin{figure}
\epsfxsize = 80 mm \epsfbox[50 180 560 660]{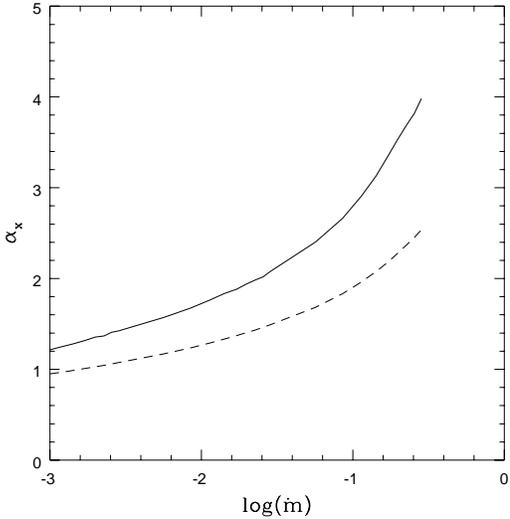}
\caption{Model A: The dependence of the 10-30 keV energy index $\alpha_x$ 
on the accretion rate for $M=10 M_{\odot}$ and the viscosity parameter 
$\alpha$ equal 0.03 (dashed line) and 0.33 (continuous line)
\label{fig:indexGBH}}
\end{figure}

Since in GBH the difference between the disc and the corona electron 
temperature is smaller the hard power law part of the spectrum is also shorter
and the spectral components are not as distinctive as in AGN. Nevertheless,
in order to show quantitatively the basic trends we plot in Figure~\ref{fig:indexGBH} the
dependence of the hard X-ray spectral index on the accretion rate and the 
viscosity parameter.  

The extension of the spectrum predicted by the model is the same as for
AGN since it is determined by the maximum value of the electron temperature
achieved in the corona (see Figure~\ref{fig:electron}).

\subsection{Model B: Monte Carlo description of Compton cooling
 and radiation spectra}
\label{sect:Monte}

Here we present the corona structure determined from the accurate description
of the Compton cooling based on the Monte Carlo computations of the 
amplification factor $A$ in Equation (8).

In this case the amplification factor depends on the soft photon energy, i.e.
the temperature distribution along the accretion disc surface. The results
of the corona structure have to be presented separately for the large masses
appropriate for AGN and small masses typical for the GBH.

Qualitatively, the radial dependencies of the corona parameters are the same
as presented in Section~\ref{sect:ModelA}. The fraction of the disc covered by the corona is
constrained to radii smaller than the critical value, $r_{max}$ and the
corona strength increases with radius for a given accretion rate till $r_{max}$
where all the energy is dissipated within a corona.

However, there are clear systematic differences between the profiles presented
in the previous Section and below.

In the case of the central black hole of $10 M_{\odot}$ the change is the
least important for high accretion rate and high viscosity. For low viscosity
and low accretion rate the change is most profound. The radial  extension of 
the corona is reduced considerably. The electron temperature is lower by a 
factor of 2. The optical depth is also slightly lower, particularly at the 
outer edge of the corona. It reflects the fact that 
the numerical value of the amplification factor is generally slightly higher 
than predicted by the Eq. (8a).

\begin{figure}
\epsfxsize = 80 mm \epsfbox[50 180 560 660]{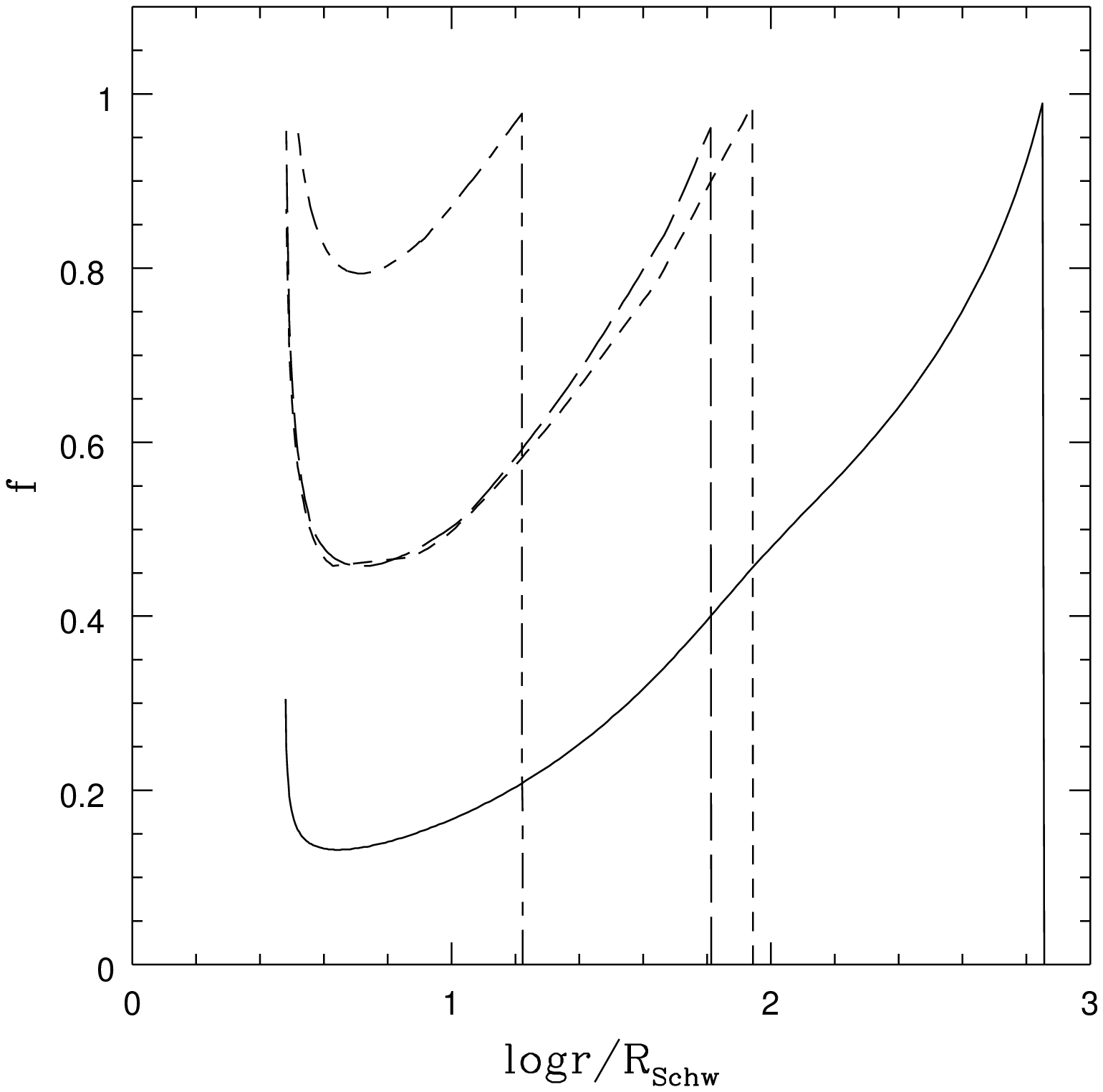}
\caption{Model B: The fraction of the energy dissipated in the corona as a function of
radius for two values of the viscosity parameter $\alpha$ and two values of
the accretion rate $\dot m$: $\dot m = 0.1$, $\alpha= 0.03$ (short dashed line),
 $\dot m = 0.1$, $\alpha= 0.33$ (continuous line),  $\dot m = 0.01$, 
$\alpha= 0.03$ (short - long 
dashed) and  $\dot m = 0.01$, $\alpha= 0.33$ (long dashed). Mass of the black hole $M = 10 M_{\odot}$. 
\label{fig:fB}}
\end{figure}

\begin{figure}
\epsfxsize = 80 mm \epsfbox[50 180 560 660]{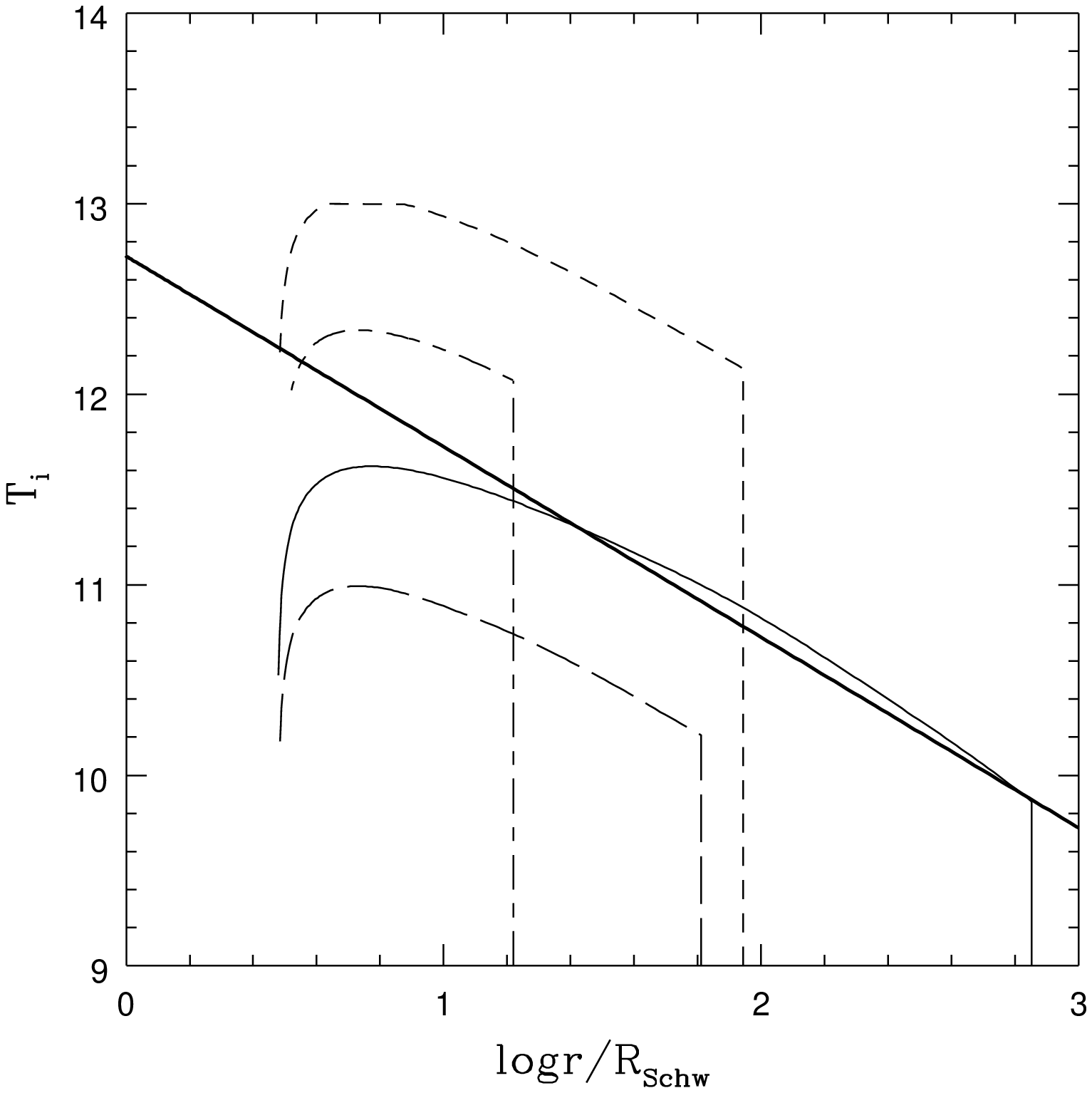}
\caption{Model B: The ion temperature as a function of
radius for two values of the viscosity parameter $\alpha$ and two values of
the accretion rate $\dot m$: $\dot m = 0.1$, $\alpha= 0.03$ (short dashed line),
 $\dot m = 0.1$, $\alpha= 0.33$ (continuous line),  $\dot m = 0.01$, 
$\alpha= 0.03$ (short - long 
dashed) and  $\dot m = 0.01$, $\alpha= 0.33$ (long dashed). The 
thick straight line shows the distribution of the local 
virial temperature. Black hole mass $M=10 M_{\odot}$.
\label{fig:ionB}}
\end{figure}

\begin{figure}
\epsfxsize = 80 mm \epsfbox[50 180 560 660]{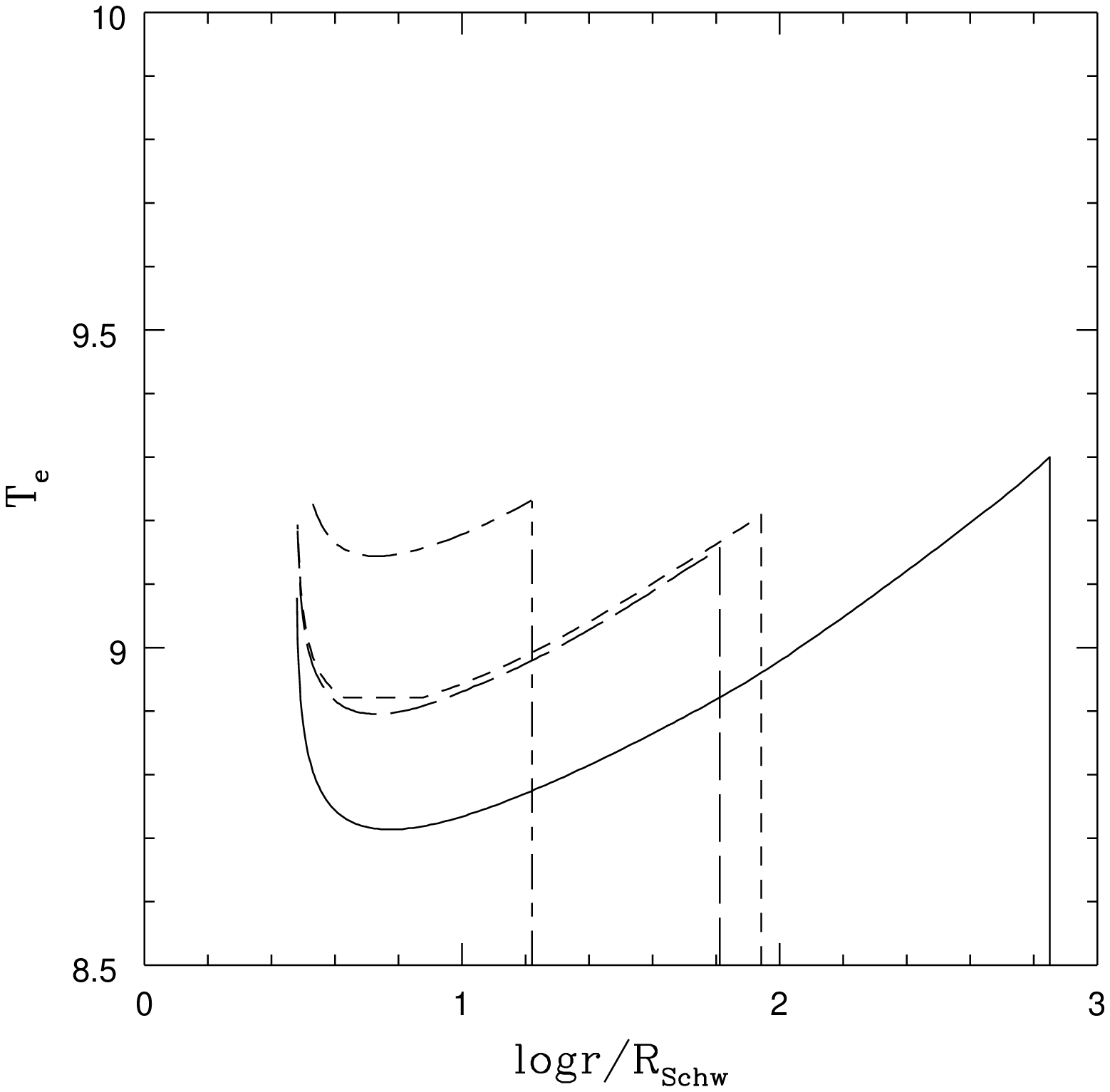}
\caption{Model B: The electron temperature as a function of
radius for two values of the viscosity parameter $\alpha$ and two values of
the accretion rate $\dot m$: $\dot m = 0.1$, $\alpha= 0.03$ (short dashed line),
 $\dot m = 0.1$, $\alpha= 0.33$ (continuous line),  $\dot m = 0.01$, 
$\alpha= 0.03$ (short - long 
dashed) and  $\dot m = 0.01$, $\alpha= 0.33$ (long dashed). Black hole mass $M=10 M_{\odot}$.
\label{fig:electronB}}
\end{figure}

\begin{figure}
\epsfxsize = 80 mm \epsfbox[50 180 560 660]{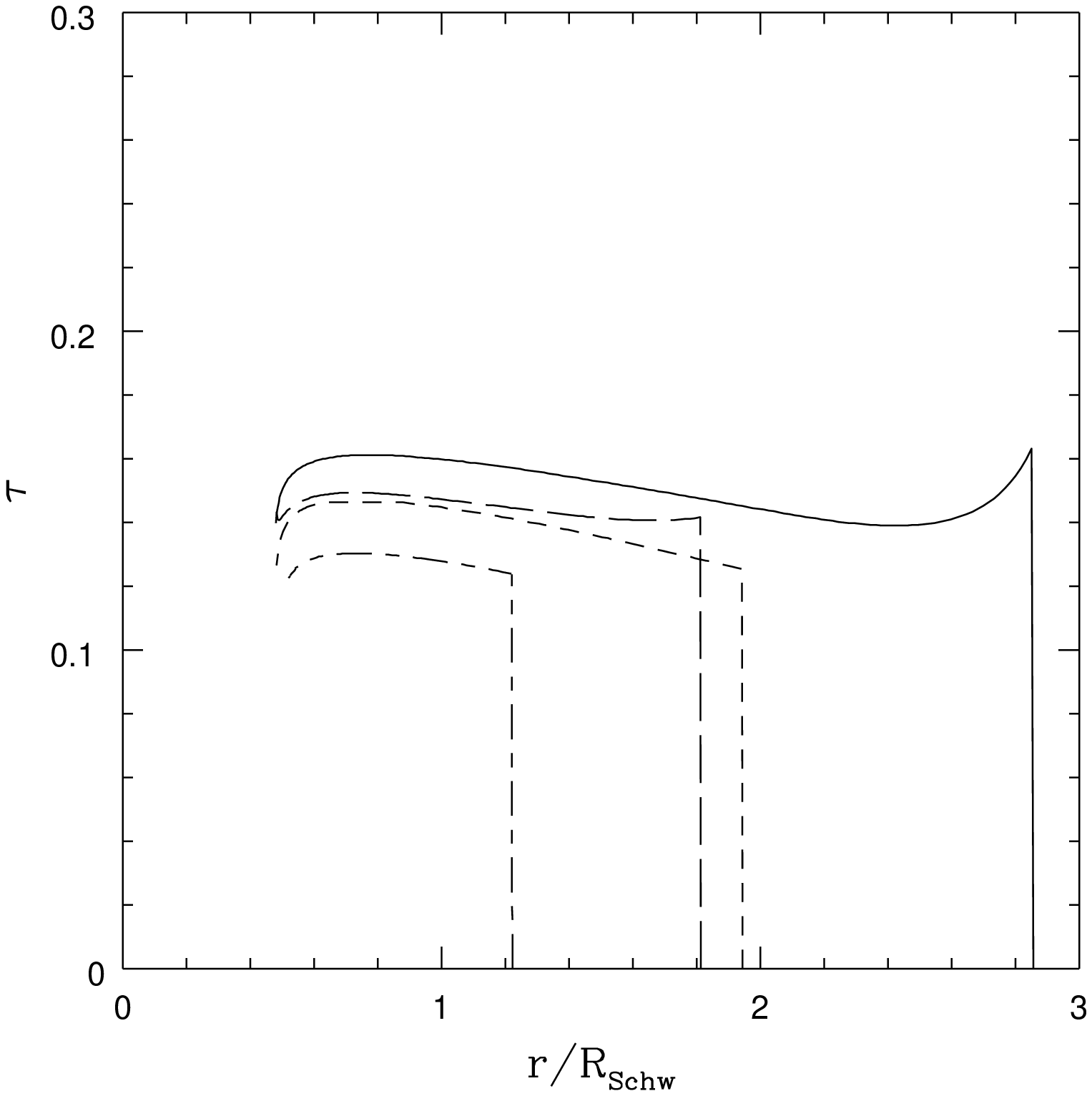}
\caption{Model B: The optical depth as a function of
radius for two values of the viscosity parameter $\alpha$ and two values of
the accretion rate $\dot m$: $\dot m = 0.1$, $\alpha= 0.03$ (short dashed line),
 $\dot m = 0.1$, $\alpha= 0.33$ (continuous line),  $\dot m = 0.01$, 
$\alpha= 0.03$ (short - long 
dashed) and  $\dot m = 0.01$, $\alpha= 0.33$ (long dashed). Mass of the black hole $M = 10 M_{\odot}$. 
\label{fig:tauB}}
\end{figure}

\begin{figure}
\epsfxsize = 80 mm \epsfbox[50 180 560 660]{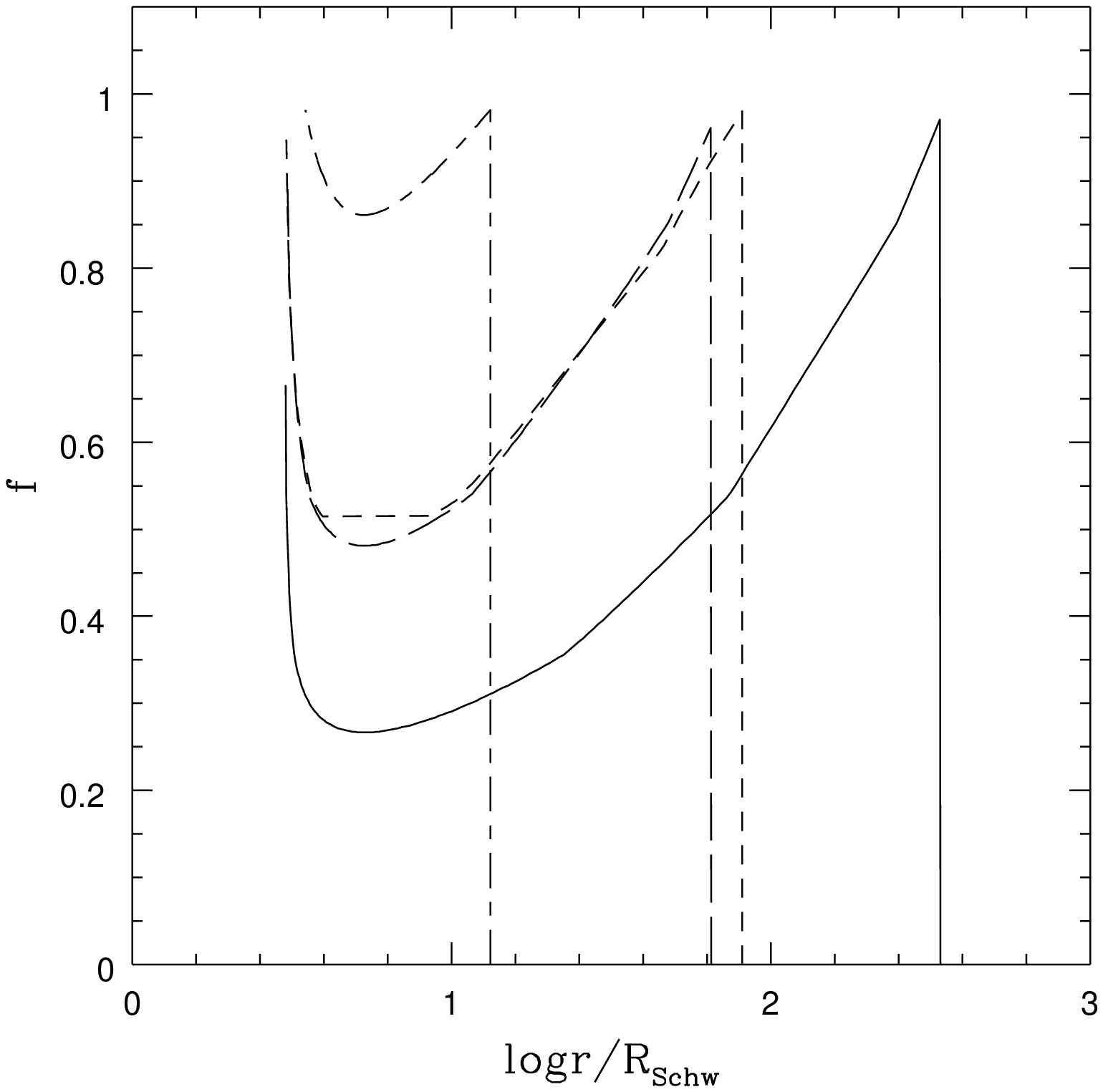}
\caption{Model B: The fraction of the energy dissipated in the corona as a function of
radius for two values of the viscosity parameter $\alpha$ and two values of
the accretion rate $\dot m$: $\dot m = 0.1$, $\alpha= 0.03$ (short dashed line),
 $\dot m = 0.1$, $\alpha= 0.33$ (continuous line),  $\dot m = 0.01$, 
$\alpha= 0.03$ (short - long 
dashed) and  $\dot m = 0.01$, $\alpha= 0.33$ (long dashed). Mass of the black hole $M = 10^8 M_{\odot}$.
\label{fig:fBAGN} }
\end{figure}

More accurate description of the Compton cooling did not help to solve the
problem of the too high ion temperature in the model. Actually, the new
results (compare Figure~\ref{fig:ion} and Figure~\ref{fig:ionB}) still enhanced the major problem of the
corona not being in hydrostatic equilibrium by giving still higher values
of the ion temperature for all presented cases of viscosity and accretion rate.

The results for AGN (i.e. assuming the mass of the black hole $10^8 M_{\odot}$)
are similar and the change from the simpler solutions presented in 
Section~\ref{sect:ModelA}
even larger. Corona shrinked also for the large accretion rate and large
viscosity and the ion temperature is equally high.

\begin{figure}
\epsfxsize = 80 mm \epsfbox[50 180 560 660]{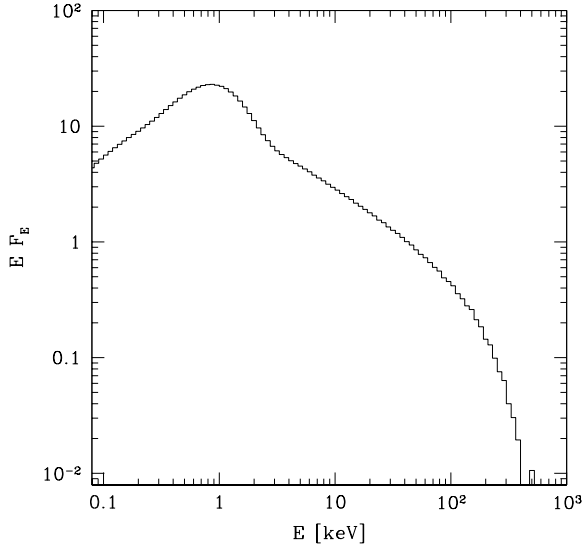}
\caption{Model B: The radiation spectrum for of the viscosity parameter $\alpha=0.33$ and
the accretion rate  $\dot m = 0.01$, 
 Mass of the black hole $M = 10 M_{\odot}$.
\label{fig:mtc_spec10}}
\end{figure}

\begin{figure}
\epsfxsize = 80 mm \epsfbox[50 180 560 660]{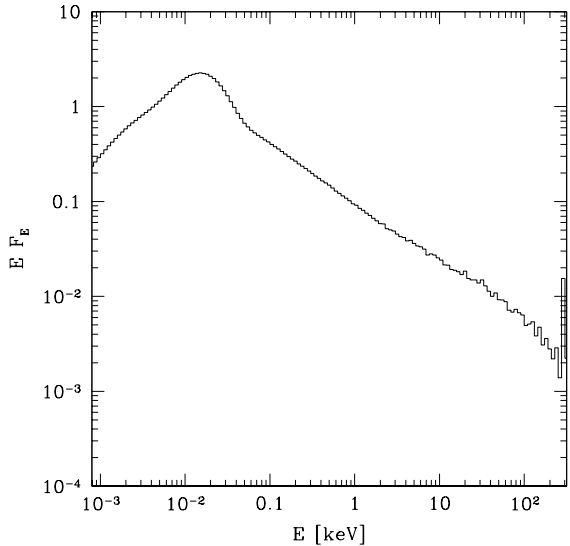}
\caption{Model B: The radiation spectrum for of the viscosity parameter $\alpha=0.33$ and
the accretion rate  $\dot m = 0.01$, 
 Mass of the black hole $M = 10^{8} M_{\odot}$.
\label{fig:mtc_spec1e8}}
\end{figure}

In Figures ~\ref{fig:mtc_spec10} and ~\ref{fig:mtc_spec1e8} we present the
radiation spectra calculated using the Monte Carlo description of Compton 
cooling. The spectral slope in case of the black hole mass equal to $10  M_{\odot}$ is almost the same as in the case of simplified description of Comptonization (Fig. ~\ref{fig:spGBH}), however in case of $10^{8} M_{\odot}$ the spectrum is much softer (Fig. ~\ref{fig:spAGN}).

\subsection{The effect of nonlocal radiative cooling}
\label{sect:nonloc}

Since most of the energy is released close to the black hole the 
radiation flux illuminating the outer parts of the disc may dominate the
radiation flux locally released due to accretion. This well known phenomenon
(see e.g. Frank, King \& Raine 1995) may have its source in either scattering of the central radiation by the extended corona or in direct illumination by the central source. The latter is important when the irradiating source is located at a certain height above a flat disc or when the disc is flaring in its outermost, gas pressure dominated region and its surface is exposed to the central flux.

The disc flaring, considered in the case of AGN, is present at radii $r>10^4 R_{Schw}$ and in our model needs not to be taken into account. 
As for the effect of both direct and scattered flux, such irradiation has been studied in some
detail for Compton heated coronae (see e.g. Kurpiewski, Kuraszkiewicz \&
Czerny 1997, and the references therein). It is important only if the luminosity of an object is very 
close to the Eddington luminosity. The same is true for our model which can be
seen from following considerations.

In the case of irradiation by the flux scattered in the corona
we can simply estimate the irradiating soft flux $F_{scat}$ 
as a function of radius
from the formula:
\begin{equation}
F_{scat}\approx {L \tau_{\perp}\over 4 \pi r^2}  \exp(-\tau_{\parallel}); ~~~~
exp(-\tau_{\parallel}) \approx (r/r_o)^{-\tau_{\perp}r \over H_P}, 
\end{equation} 
where  $\tau_{\perp}$ is the typical optical depth of the corona as measured
perpendicularly to the disc surface and $\tau_{\parallel}$ is the optical
depth of the corona integrated radially from the radius of maximum energy
generation $r_o$ up to the current radius $r$. The power law dependence
results from the radial density profile of the corona.

The direct incident flux emitted by the source located at the height 
$H_P/2$ depends on the angle between the direction of radiation and disc normal: $F_{direct}=L/4\pi D^2 \cos\Theta$, where $D\approx r$. Therefore the irradiating flux can be described by the formula:
\begin{equation}
F_{direct}\approx {L \over 8 \pi r^3}  {H_{P}\over 2} \exp(-\tau_{\parallel}). 
\end{equation}
As the locally generated soft flux depends on radius as $r^{-3}$ the ratio of the directly irradiating to the locally emitted flux depends on radius only via $\tau_{\parallel}$:
\begin{equation}
{F_{direct} \over F_{local}} \approx {1\over 12} {H_{P}\over r_{ms}} \exp(-\tau_{\parallel}(r)),
\end{equation}
where $H_{P}$ is the height at $10 R_{Schw}$, where most of energy dissipation takes place, and $r_{ms}$ denotes the radius of marginally stable orbit.
 
We computed few examples of the  model taking into account 
an increase of the soft radiation 
flux of the disc by the illumination as described above. The 
resulting electron temperature profile was hardly changed for lower accretion 
rate because of small geometrical extension of the corona and considerable
radial optical depth. The effect somewhat increased for larger values of 
accretion rate: for $\dot m = 0.1$ the temperature of the outermost 
part of the 
corona lowered as a result by 20\%, without a noticeable change in the
inner parts. The conclusion that, within the frame of our model, the
electron temperature reaches its highest value at the outer edge of the
corona is not likely to be changed if the detailed, two-dimensional radiative
transfer is solved for the corona.

\subsection{Model dependence on Coulomb coupling and ionization parameter}
\label{Coupling}

The Compton amplification factor discussed in the previous Section is one of 
the physical ingredients built into the model. Two others are: the Coulomb
coupling between the ions and electrons and the ionization parameter at the
boundary between the disc and the corona.

WCZ used the Coulomb coupling rate after Shapiro et al. (1976; see Spitzer
1962) and assumed $ln \Lambda = 20$. Results presented in 
Sections~\ref{sect:ModelA} and
~\ref{sect:Monte} followed this approach. 
This description corresponds to the pure hydrogen
plasma and it is correct only in the limit of $T_i/m_H \le T_e/m_e$. The
need for more accurate description was stressed recently by Zdziarski (1998).

We therefore replace the Equation (6) with the equation
{\setlength\arraycolsep{1pt}
\begin{eqnarray}
{dE \over dt} = - {3m_{e} \over 2m_{p}} N_{e}N_{p} \sigma_{T} c \times 
{(kT_{e}-kT_{p}) \over K_{2}(1/\Theta_{e})K_{2}(1/\Theta_{p})} \ln\Lambda\times{} 
\nonumber\\
{}\left[{2(\Theta_{e}+\Theta_{p})^2+1 \over \Theta_{e}+\Theta_{p}} 
K_{1}\left({\Theta_{e}+\Theta_{p} \over \Theta_{e}\Theta_{p}}\right)
+2K_{0}\left({\Theta_{e}+\Theta_{p} \over \Theta_{e}\Theta_{p}}\right)\right]
\end{eqnarray}}
after Stepney \& Guilbert (1983) and Zdziarski (1998), where 
$\Theta_{e}=kT_{e}/m_{e}c^2$ and $\Theta_{p}=kT_{p}/m_{p}c^2$.

The net effect of this change is a decrease of the Coulomb coupling
rate by a factor 2. The model calculated with the new  
Coulomb coupling rate for the viscosity $\alpha = 0.03$ does not
allow for any coronal solutions. Larger viscosity model
($\alpha = 0.3$) gives very similar solutions to these obtained using Shapiro
et al. description of cooling, provided the accretion rate is small 
($\dot m = 0.01$). In the case of higher accretion rates ($\dot m = 0.1$) the 
resulting optical depth of the corona is larger in its innermost parts and 
achieves value of 0.3. The corona covered region is slightly smaller in that case.

Since it was suggested (e.g. Begelman \& Chiueh 1988, Bisnovatyi-Kogan \&
Lovelace 1997) that other more efficient mechanisms
heating electrons might operate in
a two-temperature plasma we also make a simple exercise by multiplying
the original coupling rate from Equation (5) by a factor 2. This
change leads to more extended corona. For low viscosity ($\alpha = 
0.03$) the electron temperature increased up to 
$\sim 3 \times 10^9$K at the outer edge but the optical depth of the 
corona decreases down to $\sim 0.05$. The ion temperature was reduced
by a factor $\sim 3$ but it was still higher than the virial 
temperature. Further increase of the coupling rate leads to unacceptably high 
electron temperature.

The model is therefore very sensitive to the microscopic description of the
mechanism of energy transfer between ions and electrons. 

The accuracy of the description of the boundary between the disc and the 
corona reflecting the change of cooling mechanism from Inverse Compton to 
atomic cooling is more difficult to discuss. The full description of the 
disc/corona transition is not available and only partial answers to this 
problem were considered so far. Macio\l ek-Nied\' zwiecki, Krolik \&
Zdziarski (1997) studied the role of the conductivity in the corona 
characterized by direct heating of electrons (no ion-electron coupling needed). R\' o\. za\' nska \& Czerny (1996) and R\' o\. za\' nska (1999) considered 
the conductivity and both the Inverse Compton and atomic cooling but only in
the case of radiatively heated one-temperature corona.

Therefore we make a simple exercise of a change in the scaling constant 0.65
in Eq. (11). An increase of this constant to 0.71 leads to an increase of the
ion temperature thus engraving the problems with the hydrostatic equilibrium.
It also causes a slight shrink of the region occupied by the corona.
Further increase of this constant, to 0.85 causes a disappearance of the
coronal solutions for low viscosity. A decrease of this constant to 0.55
leads to slight expansion of the corona covered region, a decrease of the
ion temperature but it also decreases the optical depth of the corona.

\subsection{Corona formation}
\label{sub:outflow}

The model, as outlined in Section~\ref{sect:model}, describes a two-temperature corona in 
thermal equilibrium but does not answer the question of the 
corona formation and stability.

Thermal stability of the two-temperature plasma described as in Shapiro,
Lightman \& Eardley (1976) has been studied by Pringle (1976). 
The model has been
found to be unstable, i.e. small fluctuations of the ion 
temperature lead to the
departure from the equilibrium. The same is true about the underlying disc
when the corona is not strong enough to stabilize it (e.g. Czerny, Czerny
\& Grindlay 1986, Svenson \& Zdziarski 1994). This may pose a problem to the 
model although, on the other hand, may be convenient if we recall the fact
that all AGN and GBH are moderately or strongly variable so the stationary
equations can at best provide an adequate description of some average 
state. 

However, it also means that the presented set of equations does
not describe the physical mechanism leading to corona formation.
To answer this question we may look for instability mechanisms operating
in the bare accretion disc which may lead spontaneously to thermal
stratification and finally to the equilibrium state described by our model.

It is well known that $\alpha-$viscosity discs are unstable in their 
radiation pressure dominated parts (Pringle, Rees \& Pacholczyk 1973,
Shakura \& Sunyaev 1976). However, if a presence of magnetic field is
allowed within a disc, there is also photon bubble instability 
(Gammie 1998) which
develops in the same region and seems to be characterized by even 
shorter timescale than the thermal instability. Any of those 
instabilities may in principle lead to corona formation and the 
detailed study of this process is beyond the scope of the present paper. 

However, the attempt to identify the region of the corona development
$ r \le r_{max}$, with the onset of the radiation pressure instability
is tempting. Therefore in this section we study the radial extension
of the disc instability zone and we discuss the possibility to 
adjust either the viscosity in the disc and in the corona 
or the nonlocal outflow in such a way as to make this two regions
to coincide.

\subsubsection{Radiation pressure instability}
\label{sub:inst}

\begin{figure}
\epsfxsize = 80 mm \epsfbox[50 180 560 660]{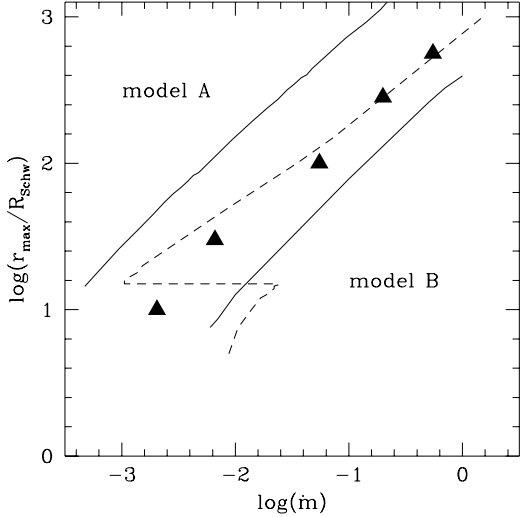}
\caption{The extension of the radiation pressure instability zone estimated
from the vertically averaged disc model (dashed line) and from the model
with the vertical structure included (triangles). The existence of inner 
stable ring for $\dot m $ between $\sim 0.025$ and $0.001$ in vertically
averaged models is caused by the opacity; this zone disappears in models
with the vertical structure. Continuous lines show the extension of
the corona covered zone given by Model A and Model B, correspondingly. 
Mass of the black hole is
$M = 10^8 M{\odot}$ and the disc viscosity parameter is $\alpha_d = 0.03$.
\label{fig:instab}}
\end{figure}

The criterion for the onset of the radiation pressure instability in a pure
hydrogen disc was given by Shakura \& Sunyaev (1976) ($\beta \le 2/5$, where
$\beta $ is the gas pressure to the total pressure ratio), and the extension 
of the region approximately corresponding to this criterion ($\beta \le 1/2$)
was already estimated by Shakura \& Sunyaev (1973).

The onset of instability, however, depends significantly on the description
of opacities (e.g. Hur\' e 1998). Applying the correct opacities to the 
vertically averaged disc structure leads to difficulties with finding the 
unstable region since complex algebraic equations allow for multiple 
solutions in the inner part of accretion disc (e.g. Hur\' e 1998 and
the references therein). Our attempt to use the vertically averaged disc 
model also resulted in peculiarities, namely, an inner stable region appeared 
in addition to the outer stable zone (see Fig.~\ref{fig:instab}).

Therefore we use the code computing the vertical structure of a disc
described by Pojma\' nski (1986)  (see also Siemiginowska, Czerny \& 
Kostyunin 1996), with new opacities (R\' o\. za\' nska et al. 1999).
We construct the $log \dot M$ vs. $log \Sigma_d$ (i.e. accretion rate
versus disc surface density in logarithmic scale) curves for several
disc radii since the negative slope of such a curve indicate a thermal
instability. 

The extension of the radiation pressure instability zone as a function of the
accretion rate for the viscosity parameter $\alpha$ in the disc equal
0.03 is shown in Figure~\ref{fig:instab}.
This zone is larger than the fraction of the
disc covered by the corona according to the model B
although it shows the same trend.  

The viscosity within a cool disc does not have to be the same as in the hot
corona. If we increase the disc viscosity or decrease the coronal viscosity
the two regions move closer to each other. For disc viscosity equal 1 the
two regions roughly coincide for larger accretion rates but for low
accretion rates the instability zone is still slightly larger than the 
corona covered region.

\section{Discussion}
\label{sect:diss}

Our disc/corona model has significant prediction power. It is only
parameterized by the mass of the black hole, the accretion rate and the
viscosity parameter $\alpha$. All measurable quantities, like the ratio
of the optically thick disc emission to the hard X-ray emission, the
soft and hard X-ray slope and the extension of the spectrum into gamma ray
band result from the model, including the trends for the change of these
quantities with the accretion rate. The model, therefore, can be tested
against the observational data.

Qualitatively, the model shows most of the trends seen in the data correctly.
According to the model, the hot plasma (corona) temperature is not uniform
but lower close to the black hole ($\sim 100$ keV) and larger further out
($\sim 300 $ keV). Such a trend coincide nicely with the determination of the
temperature structure in Cyg X-1 in a low state (Moskalenko, Collmar \& 
Sch\" onfelder 1998).  
In the case of AGN the quality of the data is not
high enough to model any departure from a single temperature fits to X-ray
data but the typical value determined from the data 
is consistent with
the radially averaged value given by our model.

The radial extension of the hot plasma ($\sim 100 - 1000 ~R_{Schw}$ in our
model) is also consistent with the size of the outer corona determined
by Moskalenko et al. (1998). It is significantly larger than in spectral data
fitting models of Cyg X-1 and other GBH (e.g. Poutanen et al. 1997) but it may be even still
too compact to satisfy the requirements imposed by the variability analysis
of Cyg X-1 (Cui et al. 1997). No similar constraints are available for AGN.

Our model predicts that the decrease of the accretion rate results in a 
relative increase in the hard X-ray emission. Therefore, it qualitatively
reproduces the time evolution of X-ray novae (Czerny, Witt \& \. Zycki 1996)
although more careful data fitting is required to confirm this conclusion
(Janiuk et al., in preparation). In the case of AGN, it reproduces well
the mean quasar spectrum and the $\alpha_{ox}$ index for most Seyfert galaxies
and quasars although there are a few objects with the Big Blue Bump more
profound than expected from the model (Czerny, Witt \& \. Zycki 1997).
Also the variability of NGC 5548 can be well represented as variations of the
accretion rate in our model (Kuraszkiewicz et al. 1997).
Our model supports the view
that Narrow Line Seyfert 1 (NLS1) galaxies are characterized by higher
luminosity to the Eddington luminosity ratio than most Seyfert galaxies
(Kuraszkiewicz et al. 1998).

Although the model is quite successful in reproducing the overall 
spectral trends in accreting black holes some problems remain.

Phenomenological models fitted to AGN and GBH indicate higher optical
depth of the corona than the values predicted by the model. The results
of Moskalenko et al. (1998) might indicate that this problem is limited
to the innermost part of the corona. It may be related to the fact
that a number of papers
suggest both on the basis of the maximum temperature achieved by the
optically thick disc and on the basis of the normalization of the 
reflection component in X-rays that the disc may not extend down to the
marginally stable orbit. It is even more clearly seen on a diagram showing
the correlation between the energy index of the 'primary hard X-ray power 
law' versus the strength of the reflection component from the disc 
(Zdziarski, Lubi\' nski \& Smith 1998). 
If the optically thick flow is actually
disrupted in the innermost part the optical depth of the hot flow
might increase. It may be caused by disc evaporation (e.g. Dullemond 1999). 

Also the prediction of the change of the coronal size with the
accretion rate may not be correct. In our model larger accretion rate 
results in expansion of the corona and relative shift of the hard
X-ray generation outwards. It (correctly) leads to a relative decrease
of hard X-ray power but at the same time predicts that the size of the
corona increases so we would expect an increase in any time delays observed
in X-rays while the opposite is true, at least in Cyg X-1 (Cui et a. 1997).
This problem may be again connected with the possibility of the disruption
of the disc in its innermost part since both the power density spectra and
time delays in Cyg X-1 indicate two component behavior.

The model is very sensitive to the description of the underlying microscopic
processes. Further work is clearly needed and at present it is difficult to 
tell whether more advanced parameterization (particularly of the disc/corona
transition) will improve the model or disqualify it.

Our model is not the only model of accretion onto a black hole. It is based on
a number of assumptions which may not be justified, mostly an assumption
that the hot phase itself also accretes and the gravitational energy of that
phase is converted into
kinetic energy of ions which in turn heat electrons through Coulomb
interaction. Models based on different assumptions can be formulated,
like advection-dominated accretion flow model (e.g. Esin et al. 1997),
accretion flow with shocks (e.g. Chakrabarti \& Titarchuk 1995, 
Chakrabarti 1997). 
Since the classical disc is thermally unstable in the radiation
pressure dominated regions a clumpy accretion disc is also a very
attractive possibility. Outlines of such models were recently discussed
by a number of authors (e.g. Collin-Souffrin et al. 1996, Krolik 1998).
However, since all models are based on assumptions difficult to support
on purely theoretical grounds the best approach is to formulate all
models in a way which allows to compare them with broad band spectral data
as well as variability properties.

\section{Conclusions}
\label{sect:conc} 

The accreting corona model reproduces qualitatively a number of properties
of AGN and GBH spectra, including the value of the typical electron 
temperature and the trends in the spectral changes with the change of the 
accretion rate. Also the initial half-qualitative comparison with the data for
a number of objects is encouraging (Kuraszkiewicz et al. 1997 for NGC 5548,
Czerny et al. 1997 for quasars and Seyfert galaxies). 

Further development 
of the model is also required. The most important task is to improve the description of the condition for the disc/corona transition in the case of two-dimensional flow. Reconsideration of this transition, taking into account both radiative and conductive heat transport, may modify the predicted fraction of energy liberated in the corona as a function of radius. It may also naturally lead to the prediction of disc evaporation:  phenomenological 
model fitting seem to indicate the necessity of such a phenomenon
(e.g. Poutanen et al. 1997, \. Zycki, Done \& Smith 1998, 
Loska \& Czerny 1997, Zdziarski et al. 1998).

\bigskip

\section*{Acknowledgements}
We thank Piotr T. \. Zycki for introducing his Monte Carlo description of
the Comptonization process into the coronal code.

\end{document}